\documentclass[11pt]{article}
\usepackage{natbib}
\usepackage{amsmath,amssymb,gensymb}
\usepackage[left=3cm,right=3cm,top=2.5cm,bottom=2.5cm]{geometry}
\usepackage{sectsty}
\usepackage{graphicx}
\usepackage{setspace}
\usepackage[dvipsnames]{xcolor}
\usepackage[font={small}]{caption}

\sectionfont{\fontsize{12}{12}\selectfont}
\subsectionfont{\fontsize{12}{12}\selectfont}

\begin{document}

\begin{center}
 {\Large{\textbf{Crater mound formation by wind erosion on Mars}}} \\ [0.5em]
 {\large{Liam Steele\textsuperscript{1}, Edwin Kite\textsuperscript{1} and Timothy Michaels\textsuperscript{2}}} \\ [0.5em]
 {\small{1. Department of Geophysical Sciences, University of Chicago, Chicago, Illinois, USA (liamsteele@uchicago.edu).}} \\
 {\small{2. SETI Institute, Mountain View, California, USA.}}
\end{center}

\onehalfspacing



\begin{abstract}
Most of Mars' ancient sedimentary rocks by volume are in wind-eroded sedimentary mounds, but the connections between mound form and wind erosion are unclear. We perform mesoscale simulations of different crater and mound morphologies to understand the formation of sedimentary mounds. As crater depth increases, slope winds produce increased erosion near the base of the crater wall, forming mounds. Peak erosion rates occur when the crater depth is $\sim$2\,km. Mound evolution depends on the size of the host crater. In smaller craters mounds preferentially erode at the top, becoming more squat, while in larger craters mounds become steeper-sided. This agrees with observations where smaller craters tend to have proportionally shorter mounds, and larger craters have mounds encircled by moats. If a large-scale sedimentary layer blankets a crater, then as the layer recedes across the crater it will erode more towards the edges of the crater, resulting in a crescent-shaped moat. When a 160\,km diameter mound-hosting crater is subject to a prevailing wind, the surface wind stress is stronger on the leeward side than on the windward side. This results in the center of the mound appearing to `march upwind' over time, and forming a `bat-wing' shape, as is observed for Mt.\ Sharp in Gale crater.
\end{abstract}

\section{Introduction}

Landscape evolution on Earth is a competition between tectonics and rainfall \citep[e.g.][]{Burbank2011}. On Mars, both these factors have been negligible for at least 3\,Gyr, allowing slow landscape evolution through aeolian processes to dominate. Thus, Mars is a natural laboratory for exploring the co-evolution of wind and landscapes \citep[e.g.][]{Holt2010, Conway2012, Brothers2013, Brothers2016}. In this study, we focus on layered sediments in craters. Most of these sediments are indurated \citep{Malin2000}, and we refer to them as sedimentary rocks.

\begin{figure*}[t]
  \begin{center}
  \noindent\includegraphics[width=1.0\textwidth]{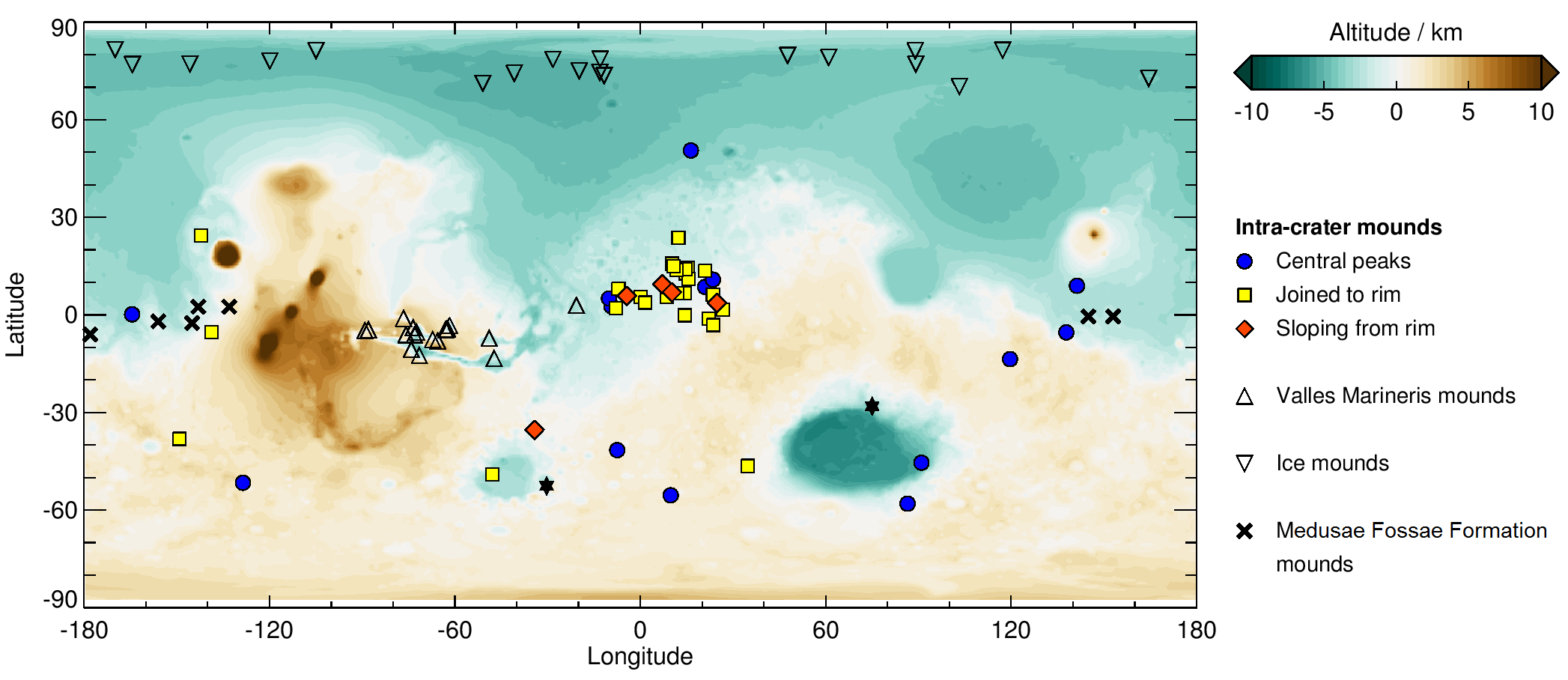}
  \caption{Global distribution of mapped sedimentary mounds on shaded MOLA topography, showing intra-crater mounds \citep{Bennett2016}, Valles Marineris mounds \citep{Kite2016}, ice mounds \citep{Conway2012} and Medusae Fossae Formation mounds \citep{Bradley2002}. The two black stars show mounds in the Terby and Galle craters that were mapped but not included in \citet{Kite2016}.}
  \label{all_mounds}
  \end{center}
\end{figure*}

\begin{figure*}[t]
  \begin{center}
  \noindent\includegraphics[width=1.0\textwidth]{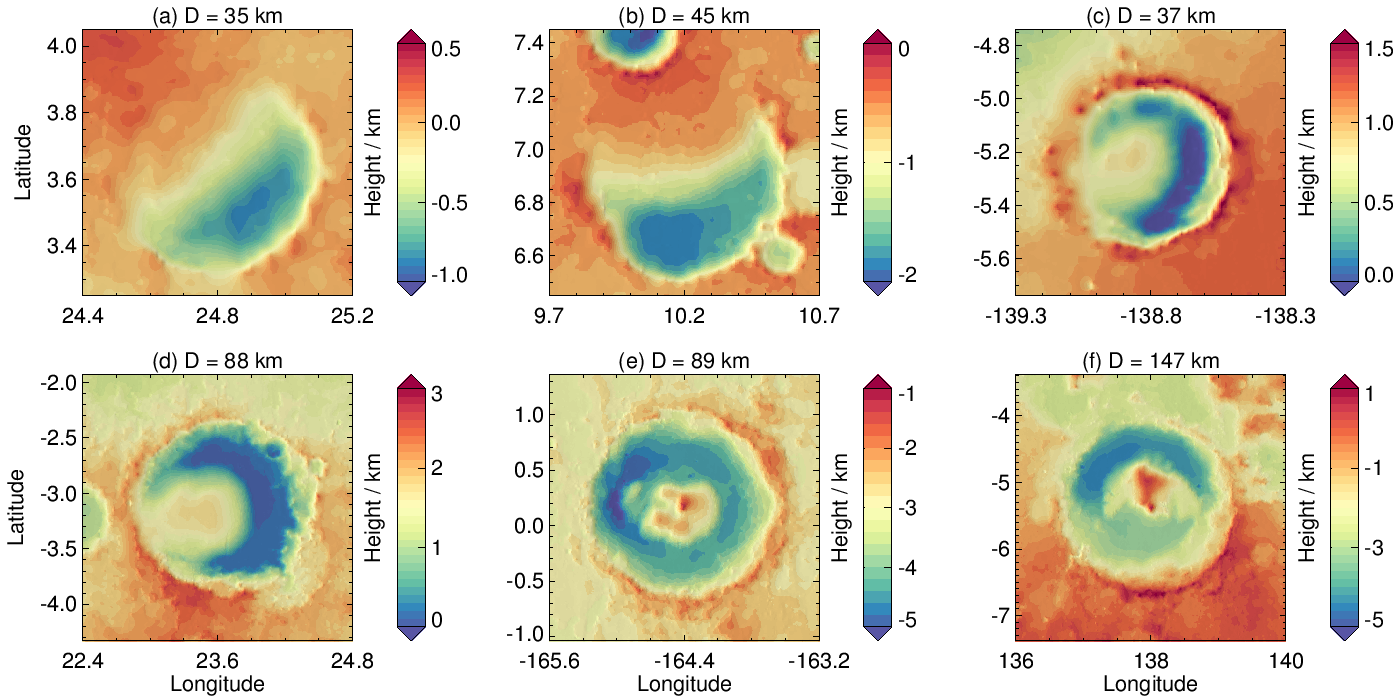}
  \caption{Mars Orbiter Laser Altimeter (MOLA) elevation data of intra-crater mounds \citep[as listed in][]{Bennett2016} showing a variety of mound morphologies, with (a,b) mounds forming a ramp down from the rim, (c,d) mounds joined to the crater wall, and (e,f) mounds encircled by moats. Craters are unnamed apart from (e) Nicholson and (f) Gale. Crater diameters are listed above each panel.}
  \label{mound_examples}
  \end{center}
\end{figure*}

\begin{figure*}[t]
  \begin{center}
  \noindent\includegraphics[scale=0.85]{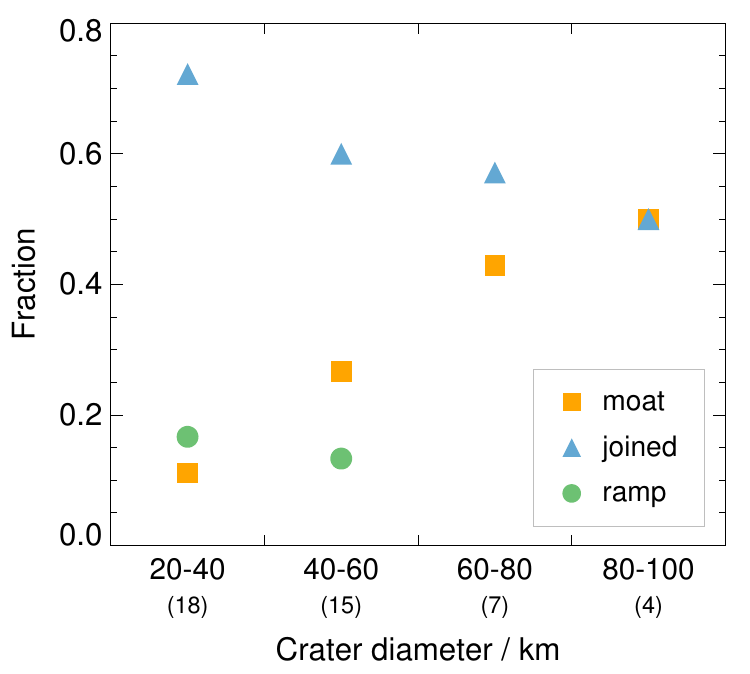}
  \caption{The fraction of craters from \citet{Bennett2016} in each size bin that have mounds displaying the characteristics of mounds encircled by moats, mounds joined to the rim or mounds forming a ramp down from the rim (see Figure~\ref{mound_examples} for mound types). The numbers in brackets under each size range show how many craters are in that bin. The Becquerel mound is ambiguous and is excluded.}
  \label{mound_types}
  \end{center}
\end{figure*}

Most of the known light-toned, post-Noachian sedimentary rocks on Mars take the form of mountains (mounds) within craters and canyons \citep{Hynek2003}, including Mt.\ Sharp in Gale crater, the target of the Mars Science Laboratory `Curiosity' rover \citep{Anderson2010, Milliken2010}. The other currently operating Mars rover, MER-B `Opportunity', is also exploring a crater that contains a sedimentary mound; the 22\,km diameter Endeavour crater \citep[e.g.][]{Squyres2012, Grant2016}. These mounds are distributed across the Martian surface, with most of the mapped intra-crater mounds located in the Arabia Terra region \citep{Malin2000, Fergason2008, Zabrusky2012, Bennett2016}. Figure~\ref{all_mounds} shows the locations of intra-crater mounds (the focus of this study), as well as mounds within the Valles Marineris canyon system \citep{Kite2016}, ice mounds in the north polar region \citep{Conway2012} and Medusa Fossae Formation mounds \citep{Bradley2002}. Visually, the intra-crater mounds mapped by \citet{Bennett2016} fall into three main types. There are mounds with a distinctive moat encircling them, mounds joined partly to the crater wall, and mounds forming a ramp down from the crater rim (see Figure~\ref{mound_examples}). The data suggest that there is a tendency for mounds completely encircled by moats (Figure~\ref{mound_examples}e,f) to become more frequent as the crater diameter increases, while mounds defined here as ramps (Figure~\ref{mound_examples}a,b) occur only in craters $<$\,60\,km in diameter (see Figure~\ref{mound_types}).

Despite the central role of mounds in the sedimentary-rock landscapes of Mars, the mechanisms responsible for mound formation and evolution remain unclear. One hypothesis for the presence of mounds is that they are the result of wind erosion of initially sediment-filled craters, with material preferentially eroded around the edges of the craters \citep{Malin2000, AndrewsHanna2010, Bennett2016}. Wind tunnel experiments carried out by \citet{Day2016} show that a mound and moat can be shaped by wind erosion, though these experiments used damp sand as opposed to sedimentary rock, with a crater model 30\,cm in diameter. Large eddy simulations \citep{Day2016, Anderson2017} suggest that vortical flows emanating from the upwind crater rim are responsible for moat excavation in sediment-filled craters, with a positive-feedback mechanism in which the erosion potential of the sediment increases the more the sediment erodes. \citet{Chan2017} present a wind-sculpted sandstone mound on Earth as an analogue to Gale crater, though it is $O(10^3)$ times smaller. Another hypothesis, motivated by outward dips in sedimentary mound strata, is that some mounds form in place by interspersed episodes of aeolian deposition and slope-wind erosion \citep{Kite2013, Kite2016}. In either hypothesis, winds play a vital role in the erosion of sedimentary deposits, and the transport of sediment within or away from the crater.

Wind erosion occurs on Mars today, as evidenced by dune field activity \citep[e.g.][]{Fenton2006, Silvestro2010, Silvestro2013, Chojnacki2011}. Observations of dune field morphologies and other aeolian features can be used to infer present-day and potential paleowind directions \citep[e.g.][]{Hobbs2010, Bridges2014, Day2016b}. Estimated sedimentary-rock erosion rates are between 0.01--50\,$\mu$m\,yr$^{-1}$, with the higher rates corresponding to vertical rock faces \citep[e.g.][]{Bridges2012, Farley2014, Golombek2014, Grindrod2014, Levy2016, Salese2016, Kite2017}. Rates $>$1\,$\mu$m\,yr$^{-1}$ allows for many kilometers of cumulative erosion \citep{Armstrong2005}. Some of the strongest winds within craters are slope winds on crater walls \citep[e.g.][]{Kite2013, Tyler2013, Tyler2015, Rafkin2016, Newman2017,Steele2017}. Due to the low density of the Martian atmosphere, the heating and cooling of the surface has a much larger impact on the near-surface atmosphere than on Earth. Due to the correspondingly strong horizontal temperature gradients, the resulting slope winds are typically 2--3 times faster than on Earth \citep[e.g.][]{Ye1990, Savijarvi1993, Tyler2002, Spiga2009, Spiga2011a}. Indeed, the strong nighttime downslope winds can increase near-surface air temperatures by up to 20\,K \citep{Spiga2011b}. 

Several processes may contribute to slope-wind erosion; (i) rock weakening and break-up by weathering and/or hydration state changes \citep[e.g.][]{Chipera2007,Wang2011}; (ii) mass wasting, followed by aeolian removal of talus to maintain steep slopes and allow continued mass wasting; (iii) aeolian erosion of weakly cemented sediments \citep{Shao2008}; and (iv) aeolian abrasion of bedrock \citep{Wang2011}. These processes range from transport-limited to detachment-limited, and predict correspondingly different shear-stress dependencies and thresholds for erosion. However, what they all have in common is the need for wind. Thus, in order to identify physical mechanisms involved in sedimentary mound formation and evolution, we need to obtain an understanding of the diurnal variation of slope winds, and the feedback between terrain evolution and circulation. To achieve this, we use a mesoscale model to simulate the circulation within craters of different morphologies. We assume detachment-limited erosion, where only the magnitude of the wind is of concern (as opposed to transport-limited erosion, where wind vectors are required for determining the transport of the eroded sediment). This is complementary to the large eddy simulations of \citet{Day2016} and \citet{Anderson2017}, where the focus was on vortical flows and not the radial slope winds.

\section{Model description} \label{sec:model}

Simulations are performed using the three-dimensional non-hydrostatic Mars Regional Atmospheric Modeling System (MRAMS) mesoscale model \citep{Rafkin2001}. This model has been used extensively to investigate many features of the Martian circulation \citep[e.g.][]{Michaels2006a, Michaels2006b, Michaels2008, PlaGarcia2016, Rafkin2002, Rafkin2003, Rafkin2009, Rafkin2016}. 

Two types of simulation were performed: `idealized' and `realistic'. The purpose of the idealized simulations is to isolate only those circulations related to crater topography. As such, the simulations have the Coriolis force and thermal tides removed, and are initialized without large-scale winds. This is similar to the approach used by \citet{Tyler2015}. Three nested grids are used, with the resolution of the innermost grid ranging between 0.5--4\,km, depending on the size of the crater being simulated (80 grid boxes span the crater diameter). There are 60 vertical levels, with the midpoint of the lowest level at 15\,m above the surface, and with 15 levels in the lowest kilometer. Tests were performed with increased numbers of vertical levels, but there were no significant changes in the strengths of the slope flows. Time steps in the outermost grid vary between 2--8\,s, depending on the crater diameter, and are reduced by a factor of two for each successive grid. The surrounding topography has constant albedo (0.23), thermal inertia (230 J\,m$^{-2}$\,K$^{-1}$\,s$^{-1/2}$) and aerodynamic surface roughness (3\,cm). For computational simplicity the transport of individual dust particles is not modeled here, and instead the visible dust optical depth at 610\,Pa is set to a constant value of 0.45. The water cycle is not included. Craters are located at 0$^\circ$N, 0$^\circ$E, at $L_\mathrm{S} = 135^\circ$, resulting in sunrise and sunset times of 05:30 and 17:30 respectively. Different times of year were tested, but the results changed little.

For the `realistic' simulations we use five nested grids, with the size of the outer grid, \textit{O}($10^4$\,km), chosen so that the crater circulations that develop on the inner grids are not directly affected by the boundary conditions. The grid spacings of the outer and inner grids are 324\,km and 4\,km respectively (decreasing by a factor of three with each successive grid). A time step of 8\,s is used in the outer grid, and this is decreased by a factor of two for each successive grid. Surface properties are interpolated from TES nighttime thermal inertia and albedo data sets \citep{Putzig2007}, with the topography from MOLA 32 pixel per degree (ppd) data \citep{Smith2001}. Output from the LMD global circulation model \citep[e.g.][]{Forget1999} is used to provide the initial conditions and boundary conditions every 1.5 Mars hours at four different times of year: $L_\mathrm{S} = 45^\circ$, 135$^\circ$, 225$^\circ$ and 315$^\circ$. For both the `idealized' and `realistic' cases, simulations are performed for 7 sols, with model data output every 20 Mars minutes. The last sol is used for analysis, in order to give the model time to `spin up', though the atmospheric temperatures and circulations patterns are repeatable after around 3--4 sols.

In this study we assume that saltation abrasion is the landscape-modifying mechanism, and that physical or chemical weathering processes break down the sediment, producing grains suitable for saltation. As such, we use the surface wind stress distributions from the simulations as a proxy for erosion. The surface wind stress is given by $\tau_* = \rho_\mathrm{a}u_*^2$, with $\rho_\mathrm{a}$ the density of the atmosphere at the surface and $u_*$ the friction velocity \citep[see][]{Kok2012}. Saltation, and hence erosion, is initiated when the wind stress is above a critical value. The saltation flux, $Q$, scales as $Q \propto \tau_\mathrm{ex}V$, where $\tau_\mathrm{ex} = \tau_* - \tau_{*\mathrm{it}}$ is the `excess' stress, $\tau_{*\mathrm{it}}$ is the impact threshold stress -- the minimum value required to sustain saltation -- and $V$ is the mean horizontal particle speed (see \citet{Kok2012} and \citet{Sullivan2017} for more details). If $V$ is assumed to increase linearly with $u_*$ then $Q \propto u_*\tau_\mathrm{ex}$, while if $V$ is constant with $u_*$ then $Q \propto \tau_\mathrm{ex}$. Previous work has assumed a linear increase of $V$ with $u_*$ \citep[e.g.][]{White1979, Armstrong2005, Almeida2008, Wang2015}, while recent work suggests the relation $Q \propto \tau_\mathrm{ex}$ should be used \citep{Sullivan2017}.

The timing of mound erosion (relative to atmospheric loss) is currently not well understood \citep{Bennett2016}, and nor is the climate at the time erosion might have occurred, as this can vary with orbital changes and atmospheric loss \citep[e.g.][]{Kite2014, Soto2015, Wordsworth2016, Ramirez2017}. Partly for these reasons, and partly because it allows us to compare our model results to reality, all simulations presented here have surface pressures similar to those of present day Mars ($\sim$6\,hPa). In general, the surface stresses predicted by our simulations are not large enough to initiate saltation, which is a situation that occurs in many other models \citep[see][]{Sullivan2017}. As such, we do not use an explicit erosion relation, and simply compare the magnitudes of the surface wind stress across the craters and mounds, relating regions of higher stress with increased potential erosion.

\section{Results}

It is not possible to simulate the entire mound formation process with a mesoscale model, but such a model can be used to take `snapshots' in time, to see how the circulation patterns would potentially erode the sediment within the crater. Initially we look at the circulation in axisymmetric craters with diameters of 40, 80 and 160\,km. This spans most of the range of mound-hosting craters cataloged by \citet{Bennett2016}, with only the smallest craters missing. The craters are surrounded by flat topography, with the results azimuthally-averaged from radial slices taken every 3$^\circ$ \citep[as in][]{Tyler2015}. Later we also look at craters covered with thick sedimentary layers, and idealized craters embedded within realistic topography. In these cases the results are not azimuthally-averaged, as the topography is not axisymmetric.

\subsection{Erosion in craters filled with sedimentary deposits}

We begin by looking at mound formation in craters containing horizontally-level sedimentary deposits. For diameters of 40, 80 and 160\,km, we assume sediment-free (basement) depths of 2.4, 3.5 and 5\,km, corresponding to data for pristine Mars craters \citep{Tornabene2018}. For each diameter, we model craters with floors that are (i) level with the surrounding flat plains, so only the crater rim protrudes, and (ii) 1.75\,km below the surrounding plains. For the 80 and 160\,km diameter craters we also consider crater floors 3.5\,km below the surrounding plains. These simulations represent different levels of sedimentary infill. We do not consider the possibility of a central peak (produced during crater formation) protruding from the sediment-filled craters. Instead we assume that if present, a central peak is below the sediment, either due to an initial small size, or through degradation \citep[e.g.][]{Robbins2012, Tornabene2014}.

\begin{figure*}[t]
  \begin{center}
  \noindent\includegraphics[width=1.0\textwidth]{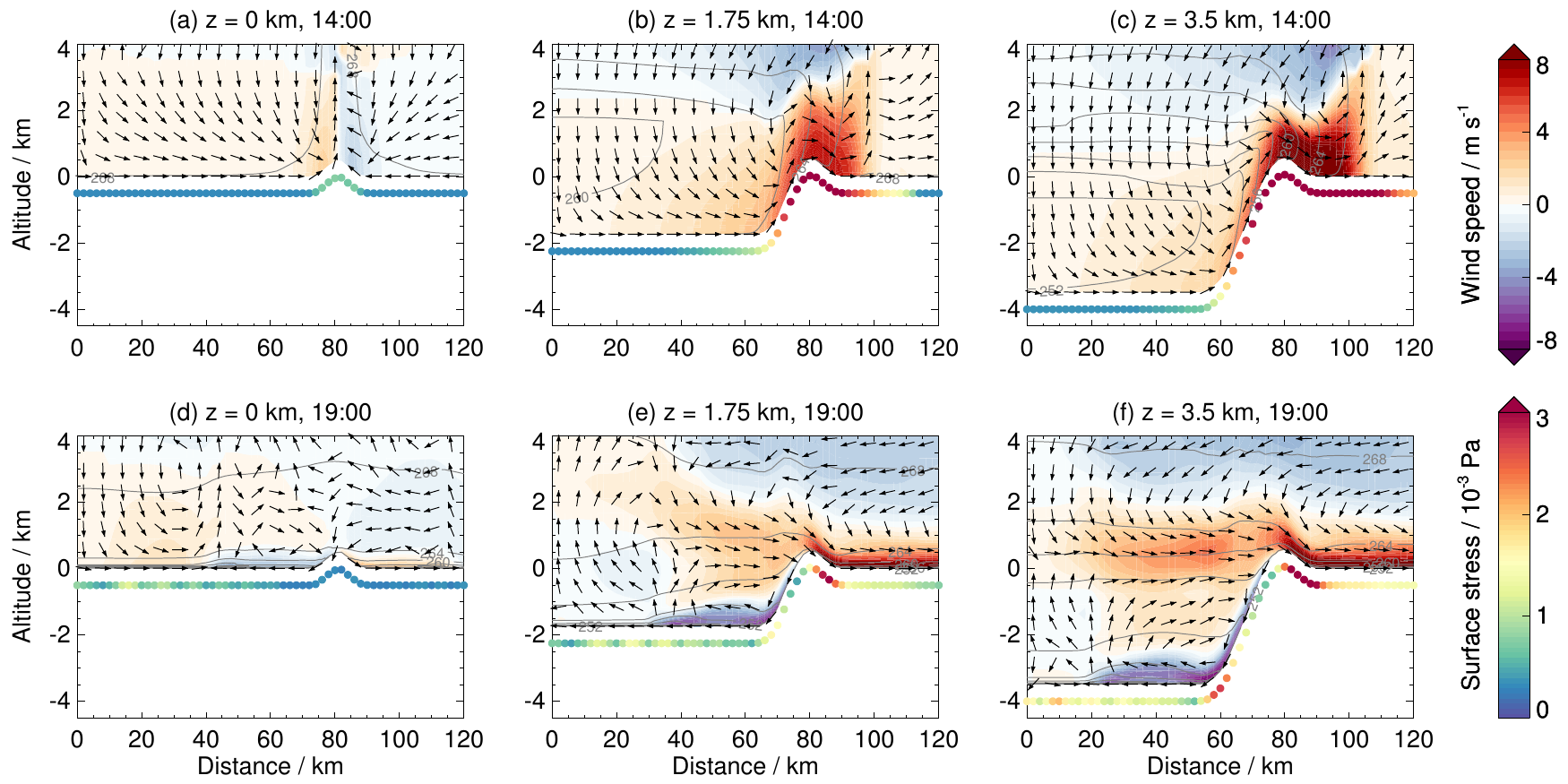}
  \caption{Azimuthally-averaged wind speed (shaded), wind direction (arrows) and potential temperature (contours) for three 160\,km diameter craters with floors at different depths, $z$, below the surrounding plains (labeled above each plot). Plots show values at 14:00 (a--c) and 19:00 (d--f) local time, with colored circles showing the maximum daytime (08:00--17:00) and nighttime (17:00--08:00) surface wind stress values in the top and bottom rows respectively. Potential temperature is contoured at 2\,K intervals in (a--c) and 4\,K intervals in (d--f).}
  \label{circulation_160}
  \end{center}
\end{figure*}

Figure~\ref{circulation_160} shows results from the 160\,km diameter simulations at (a--c) 14:00 and (d--f) 19:00 local time, as this is when the upslope winds and downslope winds (respectively) are typically at their strongest in these simulations (see Movie S1 in the supporting information for full diurnal results). Results from the smaller-diameter craters are similar, and are thus not shown. The shading shows the magnitude of the wind multiplied by the sign of the radial wind, i.e.\ $u = (u_\mathrm{r}/|u_\mathrm{r}|) \sqrt{u_\mathrm{r}^2 + u_\mathrm{v}^2}$, with $u_\mathrm{r}$ and $u_\mathrm{v}$ the radial and vertical components of the wind ($u_\mathrm{r}$ is typically an order of magnitude greater than $u_\mathrm{v}$).

\begin{figure*}[t]
  \begin{center}
  \noindent\includegraphics[width=1.0\textwidth]{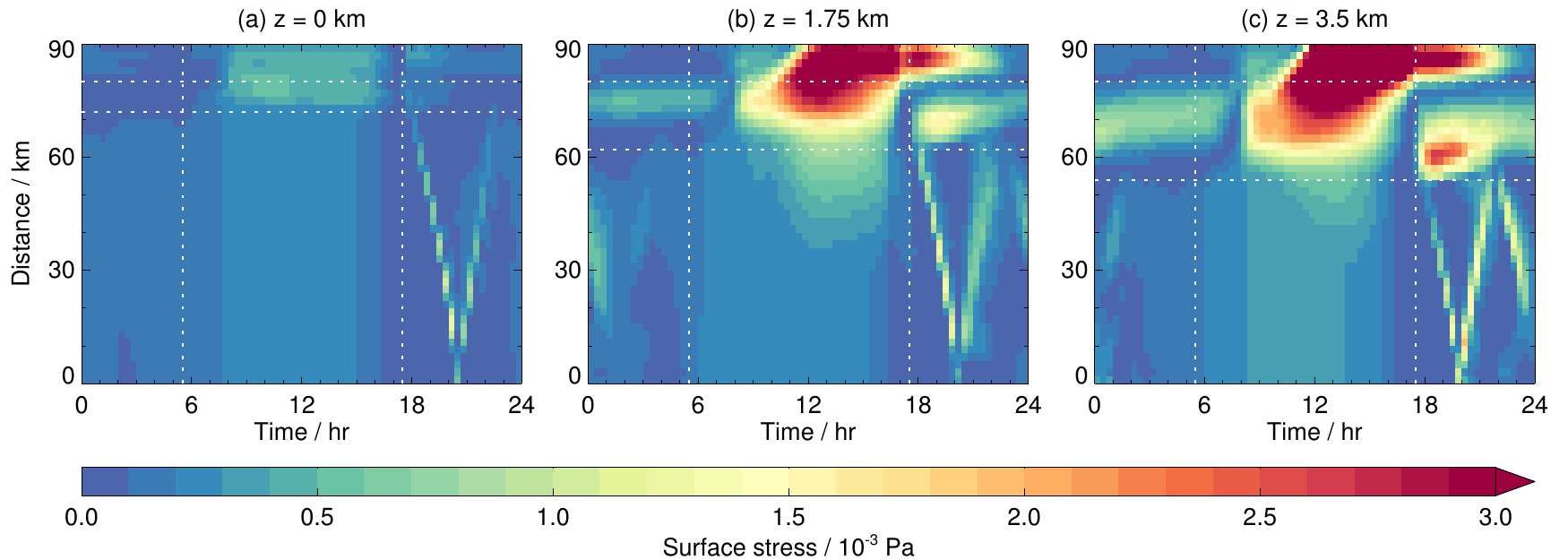}
  \caption{Surface wind stress, as a function of time of day (Mars hours) and distance from the crater center, for the three craters shown in Figure~\ref{circulation_160}. Horizontal dotted lines mark the top and bottom of the crater wall, while vertical dotted lines show the sunrise and sunset times.}
  \label{stress_160}
  \end{center}
\end{figure*}

For a given crater diameter, as the depth of the crater increases, the strength of the wind on the crater rim increases. This is caused by the larger temperature and hence pressure differences across the crater, as noted by \citet{Tyler2015}. Figure~\ref{stress_160} shows the surface wind stress for these three craters, as a function of time of day and distance from the crater center. As can be seen in Figure~\ref{stress_160}a, away from the non-erodible crater rim the largest values of surface wind stress occur on the crater floor in the evening (18:00--22:00), and are associated with air moving down the walls of the crater rim and towards the center, as seen in Figure~\ref{circulation_160}d. The `lumpy' appearance of the surface wind stress from 18:00--22:00 is a result of the discrete model output every 20 Mars minutes. For a crater filled with sediment with only the rim protruding as in Figure~\ref{stress_160}a, it is likely that passing synoptic weather systems and localized strong gusts would lead to more erosion than the nighttime downslope flow. This would likely increase the depth of the crater at all locations, though vortical flows may preferentially erode more sediment near the crater walls \citep{Day2016, Anderson2017}.

In the case of a crater 1.75\,km deep (Figures~\ref{circulation_160}b,e and \ref{stress_160}b), it is clear that on the erodible crater floor there are two daily periods of increased surface wind stress. There is again the 18:00--22:00 period associated with downslope winds which would likely erode all locations equally, but there is also now a period centered around 13:00, which is associated with upslope winds and has stress values increasing towards the crater wall. It should be noted that the stress values here result from grid box-average winds, which do not explicitly take gustiness into account. In general there is increased gustiness during the daytime \citep[e.g.][]{Fenton2010}, so peak surface wind stresses are likely to be higher than represented in the model. Even without taking this into account, erosion associated with a stress distribution like that in Figure~\ref{stress_160}b would result in more sediment being removed towards the base of the crater wall, with erosion decreasing with distance towards the center, forming a mound (assuming detachment-limited sediment transport). Erosion from traveling synoptic systems would likely be less important than for the filled case in Figure~\ref{stress_160}a, as craters get isolated from the surrounding environment as they get deeper \citep[e.g.][]{Rafkin2016}.

\begin{figure*}[t]
  \begin{center}
  \noindent\includegraphics[scale=0.8]{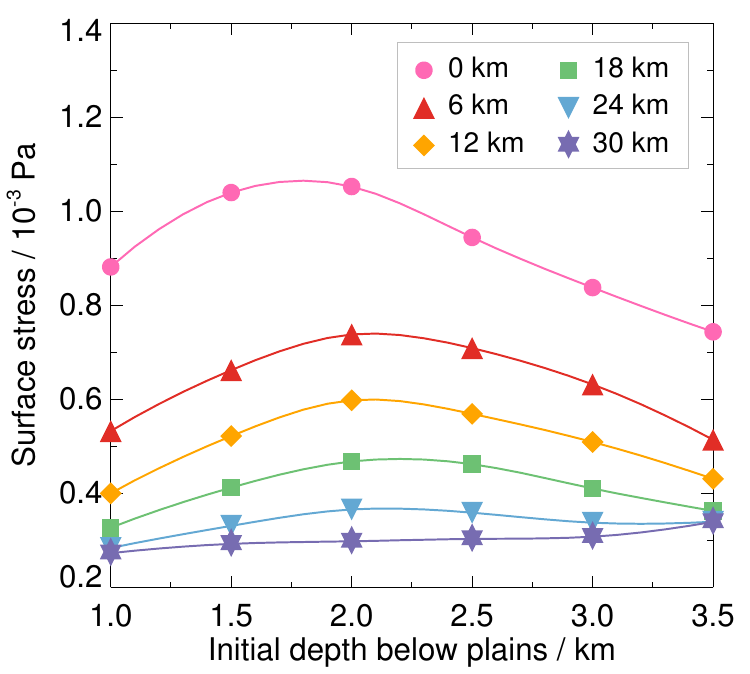}
  \caption{Variation of maximum daytime surface wind stress on the crater floor, as a function of distance from the crater wall (colored lines, ranging from 0--24\,km). Values are shown for six 160\,km diameter craters of different depths (1--3.5\,km below the surrounding plains).}
  \label{max_stress_day}
  \end{center}
\end{figure*}

At an even greater depth of 3.5\,km (Figures~\ref{circulation_160}c,f and \ref{stress_160}c), the surface wind stress associated with the nighttime downslope wind has increased, mainly because of stronger winds (as can be seen in Figure~\ref{circulation_160}f), but also due to the increasing atmospheric density in a deeper crater. However, now the stress during the daytime at the base of the crater wall is lower, and so the tendency to form a mound is reduced. Additional simulations at different depths were performed in order to understand this behavior, with the results shown in Figure~\ref{max_stress_day}. The surface wind stresses on the crater floor near the crater wall initially increase as the crater depth increases, up to $\sim$2\,km, and then start to decrease again. This maximum is a result of two competing factors. Firstly, as the crater depth increases, the daytime air over the crater at the same level as the surrounding plains gets cooler, as can be seen by the potential temperature contours in Figure~\ref{circulation_160}a--c. This results in a larger pressure difference, creating a stronger surge of air out of the crater at the rim \citep[see][]{Tyler2015}. The wind speed increases from the crater center to the rim, so initially as the depth increases the wind speeds and hence stress values on the crater floor increase. However, for crater walls of the same angle (10$^\circ$ in these simulations), as the crater gets deeper the base of the wall moves closer to the crater center (see Figure \ref{stress_160}), into a region of slower winds. These two competing factors result in the behavior seen in Figure~\ref{max_stress_day}, where a crater depth of $\sim$2\,km is the most favorable for erosion by slope winds near the crater walls.

Similar behavior is seen in the simulations with diameters of 40\,km and 80\,km. Thus, it seems plausible that a mound can begin to form by slope wind erosion if a sediment-filled crater has a depth shallower than a certain value ($\sim$2\,km in these idealized simulations). If a process results in the depth of the sediment-filled crater being much larger this value, then the reduced stress near the crater wall may result in either much slower mound formation, or possibly no mound formation at all if the threshold for sand transport is high. However, it may be possible for saltation to be initiated and maintained at lower wind speeds than the fluid threshold \citep{Kok2010, Sullivan2017}.

\begin{figure*}[t]
  \begin{center}
  \noindent\includegraphics[width=1.0\textwidth]{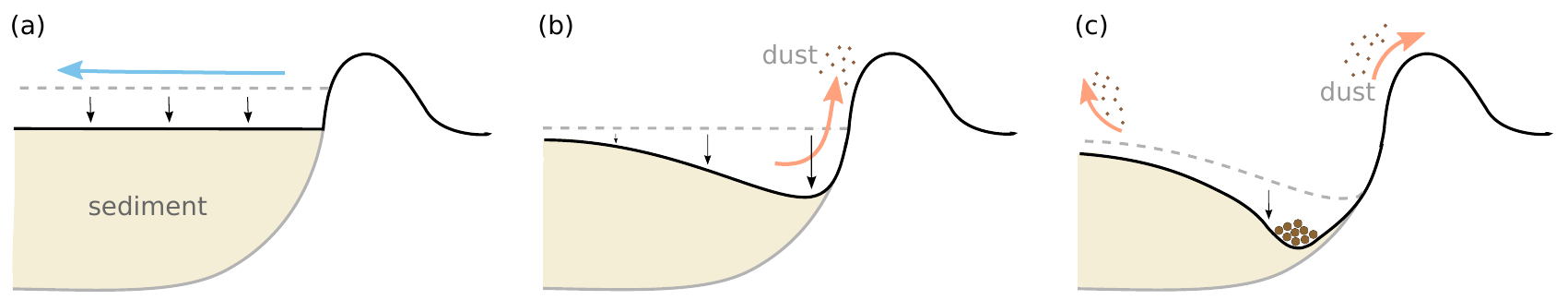}
  \caption{Cartoon showing the proposed evolution of a sediment filled crater, from initial level infill to mound formation. Red and blue arrows show the directions of the strongest daytime and nighttime winds respectively. Black arrows show how sediment erodes.}
  \label{cartoon}
  \end{center}
\end{figure*}

Figure~\ref{cartoon} shows a cartoon of our proposed method of mound formation in craters with initial flat infill. Erosion from nighttime downslope winds, as well as daytime wind gusts and dust devils which are not modeled, initially results in a gradual deepening of the crater (Figure~\ref{cartoon}a). As the crater gets deeper it gets more isolated from the surrounding environment and the upslope and downslope flows on the crater wall increase in strength, preferentially eroding sediment close to the crater wall (Figure~\ref{cartoon}b). Small dust particles can remain suspended in the air, and can be transported away from the crater. However, larger abrading clasts may accumulate at the low points on the crater floor (Figure~\ref{cartoon}c). This may increase the erosion in these areas, such as in the case of potholing in rivers on Earth \citep[e.g.][]{Pelletier2015}, leading to a positive feedback. Alternatively, a coarse-grained lag deposit can armor underlying softer rocks. Accumulations of larger particles in crater moats are observed, e.g.\ the Bagnold Dune Field in Gale crater \citep{Hobbs2010, Charles2017}.

\subsection{Mound evolution}

The results of the previous section suggest that mounds can form from craters with initial flat sedimentary infill. Thus, we next look at different mound profiles to see how they might evolve through wind erosion. Mound heights at a distance $r$ from the crater center are given by $h_\mathrm{mound}(r) = h_\mathrm{max}\cos(\pi r/2r_\mathrm{max})$, where $h_\mathrm{max}$ is the maximum height of the mound, and $r_\mathrm{max}$ is the maximum radius of the mound (which is $\le$ the crater floor radius). This profile provides a good match to the average slope of Mt.\ Sharp.

\subsubsection{Mounds in craters 1.75\,km deep}

\begin{figure*}[p]
  \begin{center}
  \noindent\includegraphics[width=1.0\textwidth]{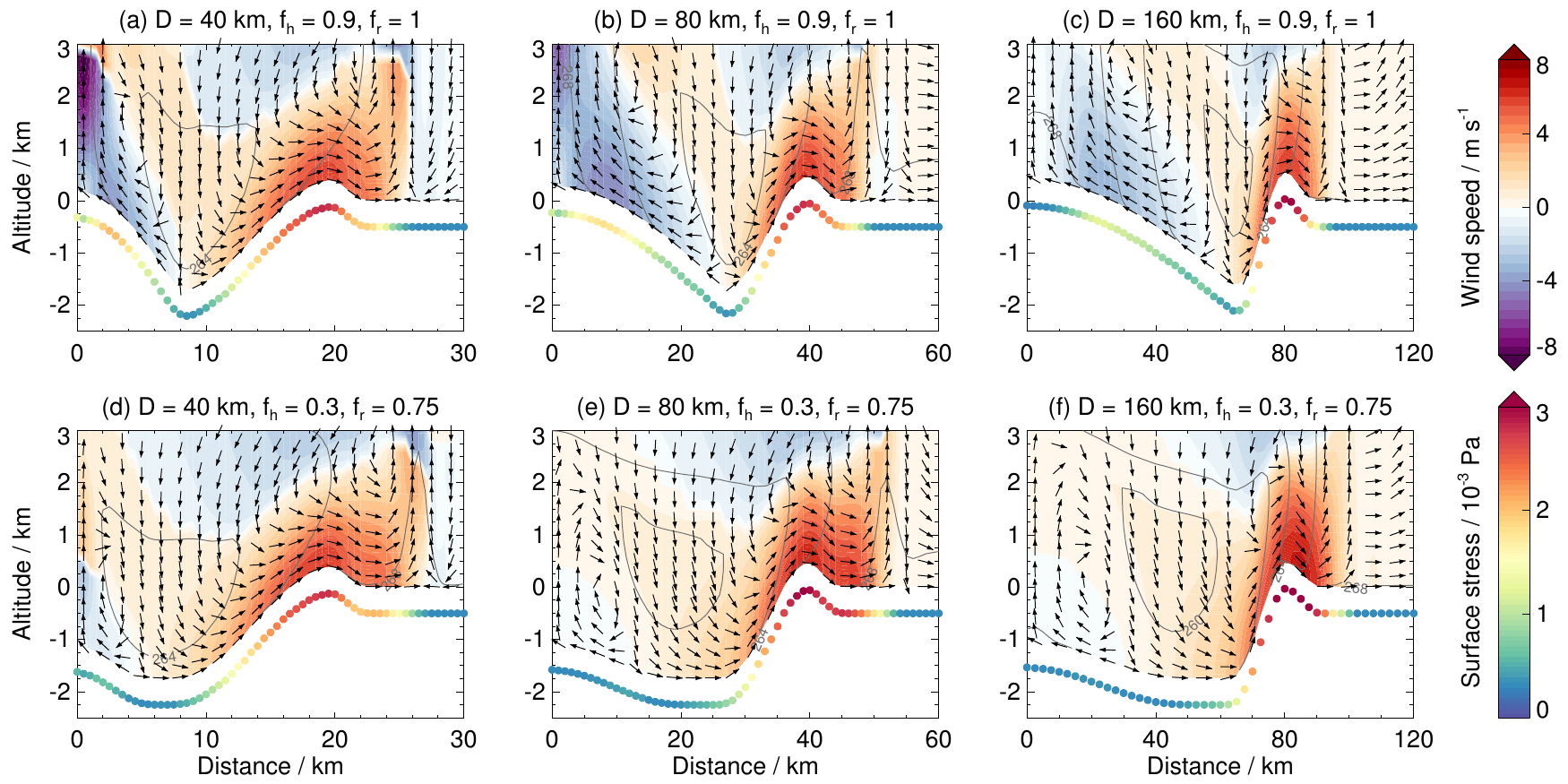}
  \caption{Azimuthally-averaged wind speed (shaded), wind direction (arrows) and potential temperature (contours) at 14:00 local time for craters with diameters of 40, 80 and 160\,km, and with mounds of different fractional heights, $f_\mathrm{h}$, and radii, $f_\mathrm{r}$, (labeled above each plot, as fractions of the crater depth and crater floor radius). Colored circles show the maximum daytime (08:00--17:00) surface wind stress values. Potential temperature is contoured at 2\,K intervals.}
  \label{wind_2pm_diff}
  \end{center}
\end{figure*}

We begin by considering two different mound shapes in craters of 40, 80 and 160\,km diameter, and 1.75\,km depth. The first mound profile begins at the base of the crater wall, and extends to 90\% of the crater depth. A mound of this type might emerge if the surface wind stress and hence erosion were maximum at the base of the crater wall and decreased towards the crater center, as suggested by Figure~\ref{stress_160}b. As the mounds are the same height but the crater diameters differ, the sides of the mound get steeper as the diameter gets smaller. The circulation patterns at 14:00, and the peak daytime stresses for these mound shapes, are shown in Figure~\ref{wind_2pm_diff}a--c (see Movie S2 in the supporting information for full diurnal results). As the crater diameter increases, the strength of the flow over the crater rim increases slightly, because of the larger temperature and hence pressure difference between the air over the crater and over the plains. Conversely, the upslope flow over the mound decreases in strength. This is because fractionally more air is lost from the smaller crater over the rim, resulting in increased downwelling, increased adiabatic warming of air in the crater, and stronger flow up the mound \citep[see][]{Tyler2015}. Due to the stronger winds blowing up the mound, the peak daytime surface wind stresses on the mound increase as crater diameter decreases. The peak stress values occur roughly 2/3 of the way up the mound in all cases. At the tops of the mounds, the peak stresses in the 40 and 80\,km craters are larger than in the moat (Figure~\ref{wind_2pm_diff}a,b), while in the 160\,km crater the peak stress at the top of the mound is lower than in the moat.

\begin{figure*}[p]
  \begin{center}
  \noindent\includegraphics[width=1.0\textwidth]{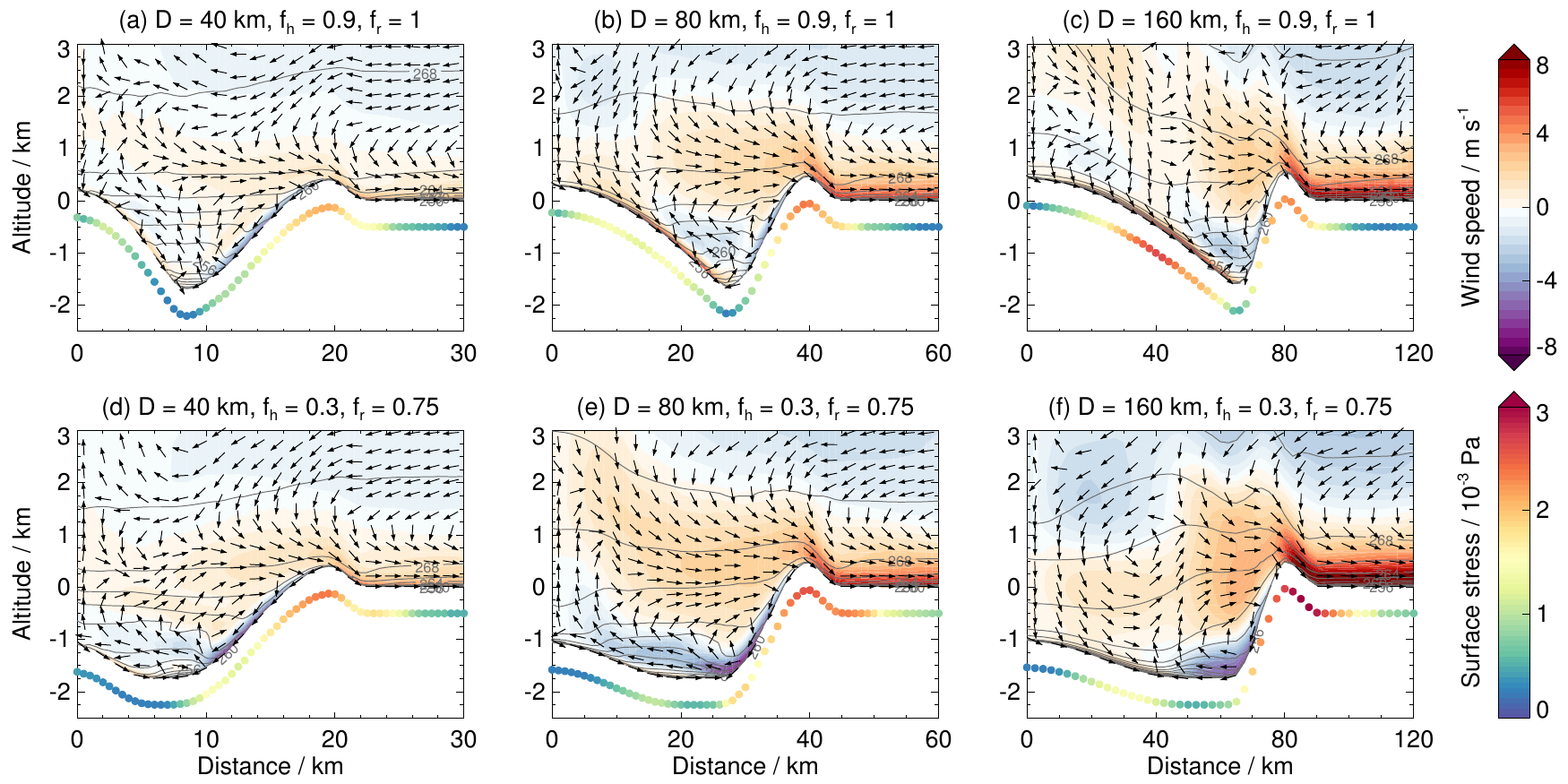}
  \caption{As Figure~\ref{wind_2pm_diff}, but for 19:00 local time, with the colored circles showing the maximum nighttime (17:00--08:00) surface wind stress values.}
  \label{wind_7pm_diff}
  \end{center}
\end{figure*}

Looking next at the circulation at 19:00, and the peak nighttime surface wind stresses (Figure~\ref{wind_7pm_diff}a--c; see also Movie S2) it can be seen that while the downslope flows on the crater walls are similar, the strength of the flow on the mound increases as the crater diameter increases. This is because the smaller crater cools more quickly, so by 19:00 the potential temperature contours are aligned horizontally, while in the larger craters the potential temperature contours are still terrain-following. This larger horizontal potential temperature gradient sustains the downslope flow for longer, resulting in larger surface wind stresses, with the peak value in the largest crater occurring about 2/3 of the way down the slope (Figure~\ref{wind_7pm_diff}c). In the smallest crater, the stresses are again larger towards the top of the mound compared to the moat (Figure~\ref{wind_7pm_diff}a). Daytime and nighttime stress distributions such as these suggest that, if all other factors were held equal, the mounds in the 40 and 80\,km craters would likely erode more at their tops than at their bases, eventually becoming more squat, while the mound in the 160\,km crater would erode more at the sides and base than at the top, becoming steeper-sided. This may be one of the reasons why larger diameter craters have a greater frequency of mounds surrounded by moats (see Figure~\ref{mound_types}).

\begin{figure*}[t]
  \begin{center}
  \noindent\includegraphics[width=1.0\textwidth]{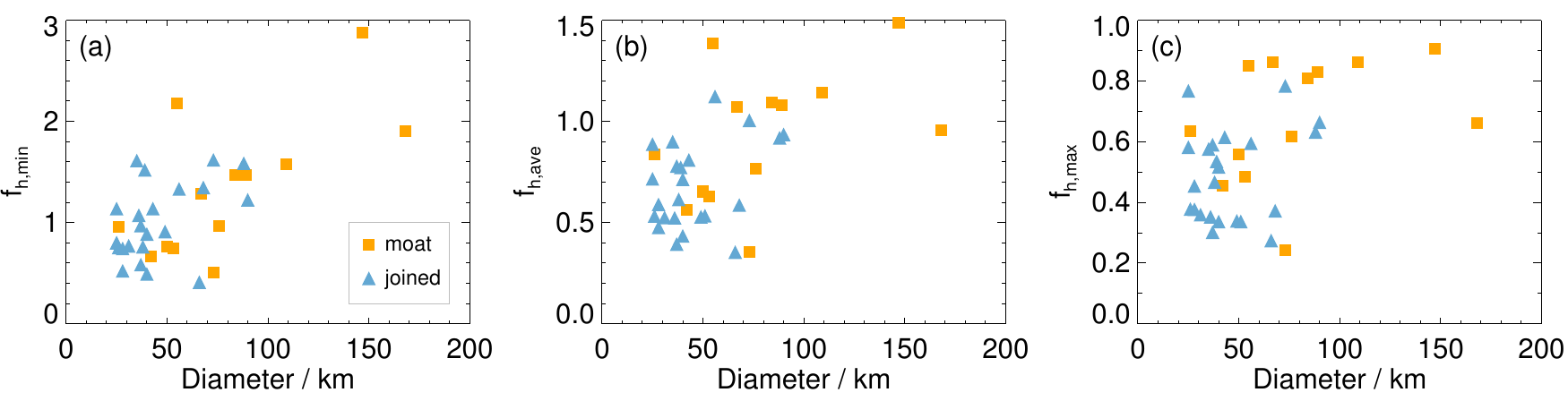}
  \caption{Mound heights expressed as a fraction of the distance between the crater floor and (a) the minimum, (b) the average, and (c) the maximum rim height. Data are for mound-hosting craters listed by \citet{Bennett2016}, with the mound and rim heights determined from MOLA 128 ppd data (errors are smaller than the symbol sizes).}
  \label{mound_heights}
  \end{center}
\end{figure*}

Figure~\ref{mound_heights} shows how mound heights in craters identified by \citet{Bennett2016} compare to their host craters. Heights for each crater were determined by taking radial slices through MOLA 128 ppd data every 0.5$^\circ$, and then calculating the maximum crater depth, the maximum mound height (ignoring central uplift peaks from crater formation), and the minimum, maximum and average rim heights. It can be seen that there is a tendency for smaller craters to have proportionally smaller mounds, suggesting more erosion of the mound tops. Indeed, from $10^5$ Monte Carlo bootstrap trials fitting a linear trend line to Figure~\ref{mound_heights}c, in only 36 cases did a negative slope result. This behavior is in agreement with the stress distributions, which show mounds in smaller craters experience greater surface wind stresses towards the tops of the mounds than do larger craters.

We now consider a much smaller mound that extends to 75\% of the radius of the crater floor, and 30\% of the crater depth. This morphology may be the result of erosion over a long time scale in a crater initially filled with sediment, or early erosion in a crater only filled with a small layer of sediment. During the daytime (Figure~\ref{wind_2pm_diff}d--f; see also Movie S2) the upslope winds along the crater walls and out over the rim are slightly stronger than for the case with the larger mound (Figure~\ref{wind_2pm_diff}a--c). This is because a smaller mound results in a greater volume of air within the crater, and a greater average distance between the crater floor/mound (which heats up rapidly during the day) and the air that is level with the surrounding plains. As such, the air is cooler by $\sim$2--3\,K, and hence there is a larger pressure difference driving the outward surge of air away from the crater. 

As the smaller mounds sit lower in the crater, they are affected more by the downwelling and associated adiabatic warming discussed earlier, and hence near-surface temperatures are warmer than for the larger mounds. At the tops of the mounds, the temperatures are $\sim$5\,K warmer in the 40\,km diameter crater, and $\sim$2\,K warmer in the 80 and 160\,km diameter craters. However, the stronger downwelling and denser atmosphere results in weaker upslope winds on the mound flanks, and so surface wind stresses are lower ($\sim$20--50\% of the values on the larger mounds). For the 40\,km diameter crater the stress is still larger near the top of the mound than at the base, while in the 80 and 160\,km diameter craters the stress values are similar along the mounds.

By 19:00 (Figure~\ref{wind_7pm_diff}d--f; see also Movie S2), near-surface temperatures over the mounds have cooled by around 10\,K, and downslope winds are at their strongest. Surface wind stresses on the mound in the smallest crater show little variation, with a slight increase towards the top of the mound. In the 80 and 160\,km diameter craters, the greater potential temperature difference near the mound flanks means downslope winds can exist for longer (as discussed earlier), resulting in stress values that increase towards the mound base. (The surface wind stresses are again $\sim$20--50\% of the values on the larger mounds.) These results suggest that as the mounds become more eroded and exist deeper within the craters, the weaker near-surface circulation produces less erosion. This may explain why intra-crater mounds persist today, rather than wind erosion removing them completely.

\subsubsection{Mounds in craters 3.5\,km deep}

\begin{figure*}[t]
  \begin{center}
  \noindent\includegraphics[width=1.0\textwidth]{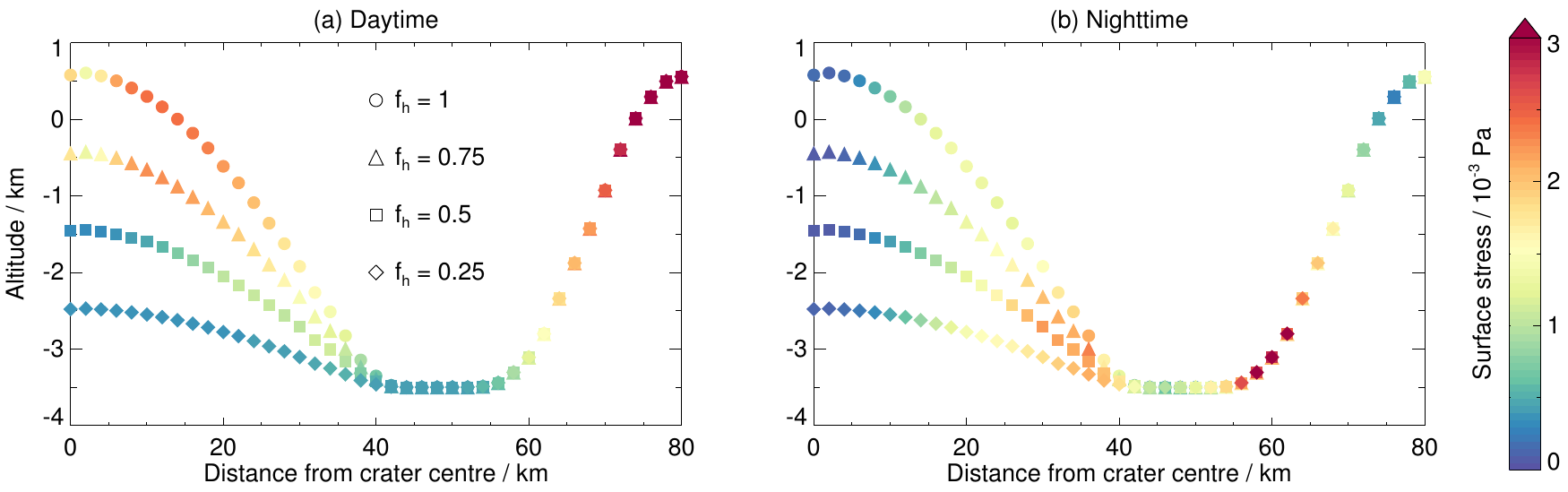}
  \caption{Maximum daytime and nighttime surface wind stress (colors) from simulations of craters 160\,km in diameter and 3.5\,km deep. The symbol locations show the mound profiles, where $f_\mathrm{h}$ denotes the fractional height of the mound in relation to the crater depth (the fractional radius is 0.7 in all cases).}
  \label{max_stress_160}
  \end{center}
\end{figure*}

We now briefly look at mounds in craters where the moat is 3.5\,km below the surrounding plains, focusing on craters with diameters of 160\,km. We consider mounds with radii extending to 70\% of the crater floor radius, and heights ranging from 25\% to 100\% of the crater depth. The maximum daytime and nighttime surface wind stresses are shown in Figure~\ref{max_stress_160}. The circulation patterns are not shown, but follow the behavior seen in Figures~\ref{wind_2pm_diff}c,f and \ref{wind_7pm_diff}c,f.

At the crater rim the temperature fields are nearly identical in the different simulations, and thus the daytime surges of air away from the craters are similar. Near-surface daytime temperatures increase as the mound height decreases (through adiabatic warming associated with downwelling) with temperatures over the shortest mound ($f_\mathrm{h} = 0.25$) being $\sim$5\,K warmer than over the tallest mound ($f_\mathrm{h} = 1$). As for the 1.75\,km deep craters, the upslope winds decrease in strength as the mounds become shorter, due to the combination of increased air density and downwelling. As such, the surface wind stresses also decrease (Figure~\ref{max_stress_160}a). In the 160\,km diameter, 1.75\,km deep crater, the stress at the top of the mound was the lowest of any point within the crater (see Figure~\ref{wind_2pm_diff}c). This is not true for the tallest two mounds in the 3.5\,km deep case, where stronger winds at the mound tops produce stresses larger than in the moat.

At nighttime, the acceleration of the downslope winds causes the surface wind stress to increase towards the base of the mounds (Figure~\ref{max_stress_160}b). For the mounds with $f_\mathrm{h} = 0.25$ and $f_\mathrm{h} = 0.5$, the nighttime stress values at the mound base are the largest of any time of day. Such mounds may be expected to erode more on the flanks and become steeper-sided. For the mound with $f_\mathrm{h} = 1$, the maximum daytime stress from upslope winds is larger than the maximum nighttime value. Thus, this mound may be expected to erode more at the top, becoming more squat. If this were to occur, it may eventually reach a height where the stress was larger towards the base of the mound, and it would then erode more horizontally. (For the mound with $f_\mathrm{h} = 0.75$, the maximum daytime and nighttime stress values are about equal.)

\subsection{Erosion in craters covered with sedimentary deposits}

\begin{figure*}[t]
  \begin{center}
  \noindent\includegraphics[width=1.0\textwidth]{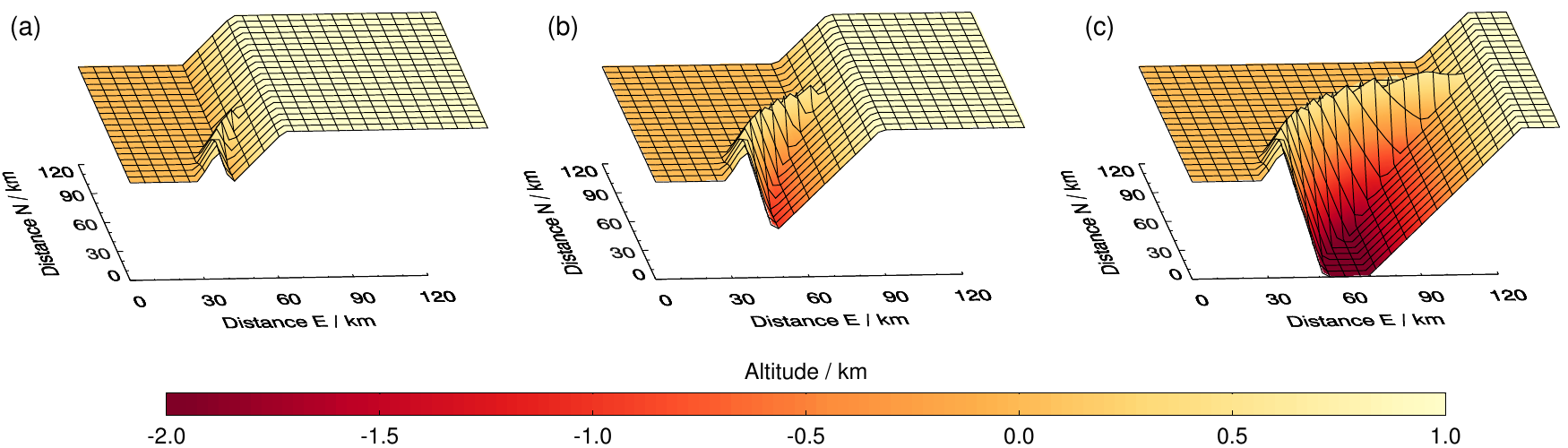}
  \caption{Morphologies of three 160\,km diameter craters covered to different extents by a 1\,km deep sedimentary layer. This represents the retreat of the layer over time from west to east. (Time evolution between the three morphologies is not considered.)}
  \label{sedim_topog}
  \end{center}
\end{figure*}

\begin{figure*}[t]
  \begin{center}
  \noindent\includegraphics[width=1.0\textwidth]{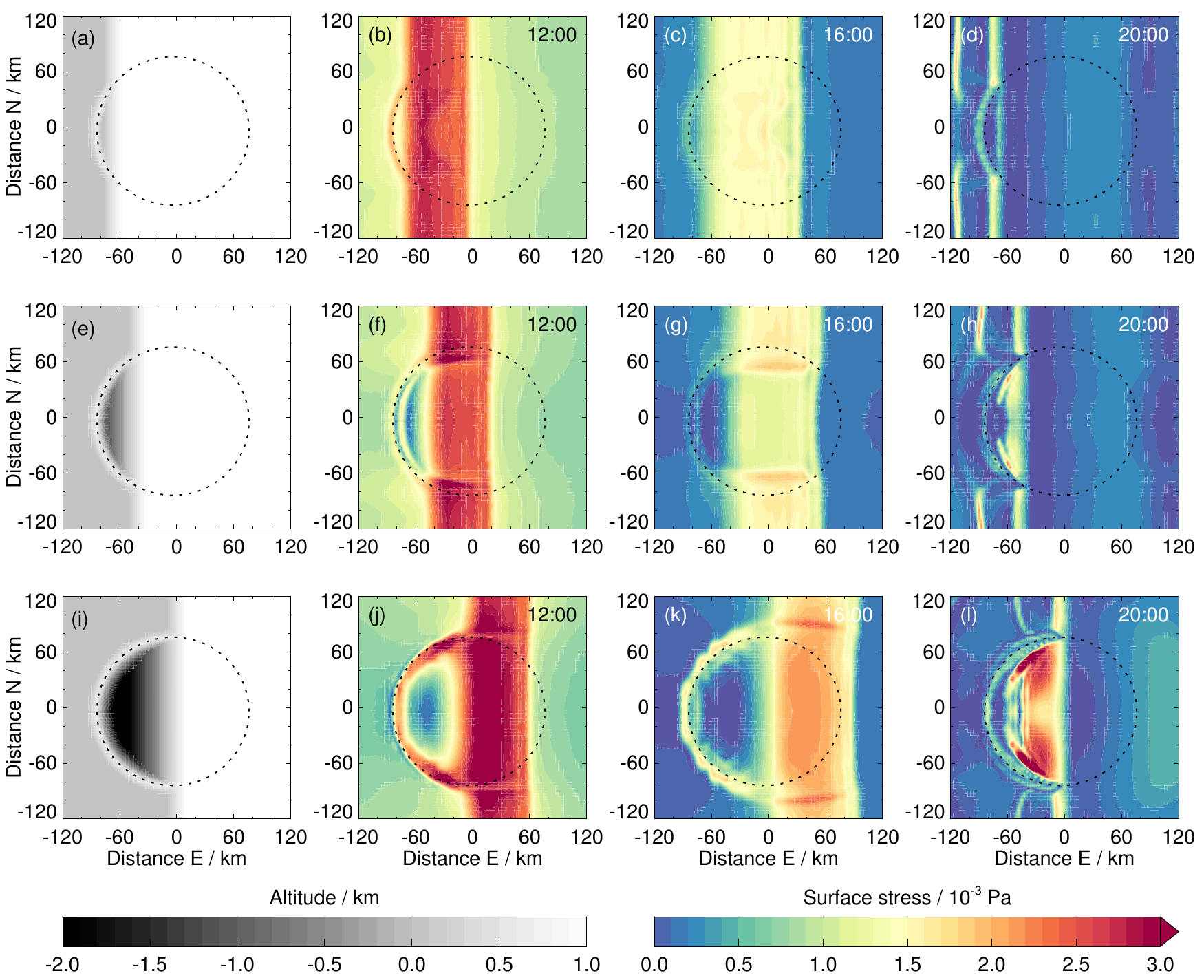}
  \caption{Topography (left column) and surface wind stress values at three different times (remaining columns) for three 160\,km diameter craters covered to different extents by a 1\,km thick sedimentary layer (each row shows a different crater morphology). The dotted line shows the location of the crater rim, and the local times are labeled on each panel.}
  \label{sedim_layer}
  \end{center}
\end{figure*}

So far we have considered erosion in axisymmetric craters filled with sediment and surrounded by flat topography. However, it has been suggested that some intra-crater mounds are the result of the erosion of large sedimentary deposits that existed on top of craters, particularly for mounds in the Arabia Terra region \citep{Fergason2008, Bennett2016}. We have therefore performed simulations of 40, 80 and 160\,km diameter craters with a 1\,km thick sedimentary layer partially covering the craters to different extents (to represent the gradual erosion and retreat of the layer over time). Figure~\ref{sedim_topog} shows slices through the 160\,km diameter craters to highlight the morphologies, and Figure~\ref{sedim_layer} shows results for the same craters. In these simulations the sedimentary layer slopes in the east-west direction, with no variation in the north-south direction. The gradient is 0.06 (slope angle $\sim$3.4$^\circ$), which was chosen to be similar to the gradients of the sloping mounds seen in Figure~\ref{mound_examples}a,b. Three nested grids are used, with an inner grid spacing of 5\,km.

When the sedimentary layer covers almost the whole crater, leaving only the rim exposed, the meridional wind has little effect on the circulation, and it is the zonal wind blowing up and down the face of the sedimentary layer that causes the surface wind stress distributions (Figure~\ref{sedim_layer}a--d). As such, the layer would be expected recede to the east uniformly over time. We assume that as the layer recedes the sediment within the crater will also be eroded, deepening the crater. As such, we next model a crater where the sedimentary layer covers 3/4 of the crater diameter, and the crater depth has increased (Figure~\ref{sedim_layer}e--h). In this case, the daytime flow up the face of the layer is still generally uniform across the crater, but there is increased surface wind stress on top of the layer near the crater rim due to the zonal wind being funneled by the topography. During the evening and night, the downslope flow along the crater walls and sedimentary layer results in increased surface wind stresses towards the crater walls (Figure~\ref{sedim_layer}h). Similar behavior is seen for the case where the sedimentary layer covers half the crater (Figure~\ref{sedim_layer}i--l), with the deeper crater allowing for stronger downslope winds and increased stresses towards the crater walls (Figure~\ref{sedim_layer}l).

The surface wind stress patterns in these simulations suggest that as a sedimentary layer recedes across a crater, it will erode more at the edges of the crater, resulting in a crescent-shaped moat. The behavior shown for these 160\,km diameter craters also occurs for 80\,km diameter craters, but in the 40\,km diameter simulations the behavior suggests just a linear retreat of the sedimentary layer, with no clear signal for the formation of a crescent-shaped moat. Wind tunnel experiments by \citet{Day2016} showed that a crescent-shaped moat can form if there is a uni-directional wind blowing across a crater. However, their experiments used 30\,cm and 60\,cm diameter crater models, and so the mound at all times is impacted by the prevailing wind. In large diameter craters, the mound could be many tens of kilometers away from the crater rim, lessening the impact of the large-scale wind blowing across the crater. However, large eddy simulations suggest that vortical flows can also result in crescent-shaped moats \citep{Day2016, Anderson2017}. Thus, in smaller craters ($<40$\,km diameter) vortical flows can explain crescent-shaped moats, while in larger craters both vortical flows or the erosion of a covering sedimentary layer by slope winds are possible mechanisms.

\subsection{Erosion in a realistic atmosphere}

\begin{figure*}[t]
  \begin{center}
  \noindent\includegraphics[width=1.0\textwidth]{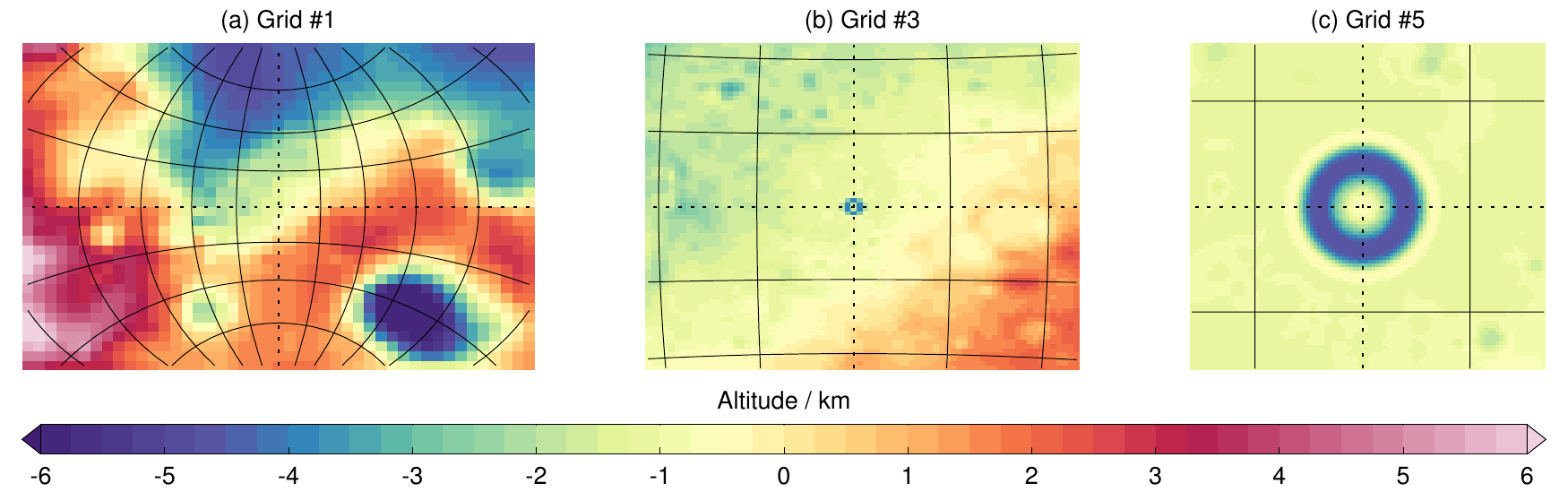}
  \caption{Topography on grids 1, 3 and 5 of the `realistic' simulations (resolution 324, 36 and 4\,km respectively). Black contours show the latitude and longitude in intervals of 10$^\circ$, 5$^\circ$ and 2$^\circ$ respectively. Dotted lines show 0$^\circ$N, 0$^\circ$E.}
  \label{real_nests}
  \end{center}
\end{figure*}

\begin{figure*}[t]
  \begin{center}
  \noindent\includegraphics[width=0.85\textwidth]{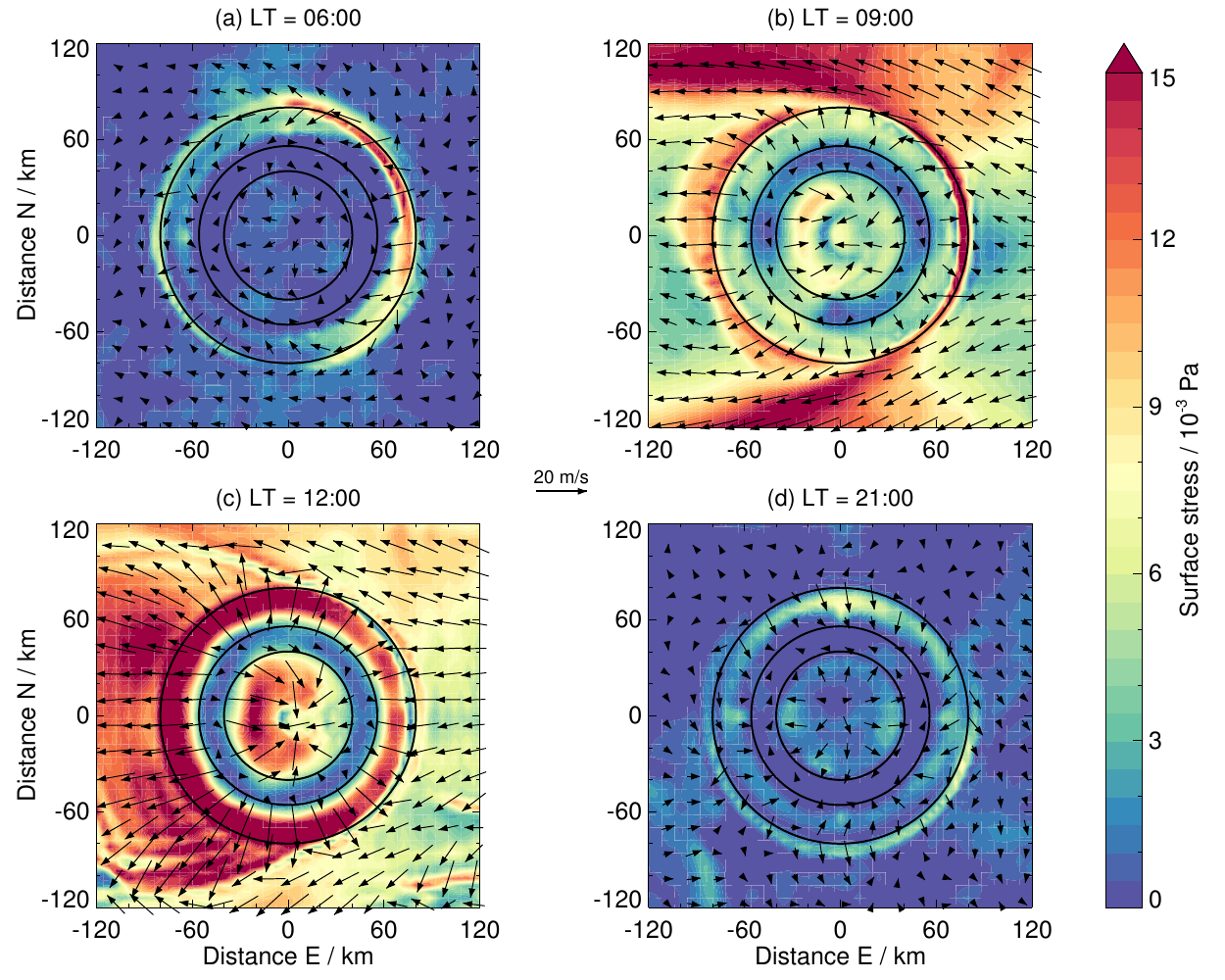}
  \caption{Surface wind stress (shading) at four different local times from a simulation of a 160\,km diameter and 3.5\,km deep axisymmetric crater at $L_\mathrm{S} = 315^\circ$. Arrows show the wind speed and direction, while the three black circles denote the locations of the crater rim and the bases of the crater and mound walls.}
  \label{real_atmos}
  \end{center}
\end{figure*}

The simulations performed so far lack the Coriolis force, thermal tides, initial large-scale winds, and realistic topography. This was in order to isolate the topography-windfield coupling. To compare these idealized simulations with reality, we performed additional `realistic' simulations using GCM boundary conditions (see section \ref{sec:model} for a description of the method). An idealized axisymmetric crater (160\,km in diameter and 3.5\,km deep, with a mound covering 70\% of the crater floor radius and the full depth of the crater) was placed at 0$^\circ$N, 0$^\circ$E (see Figure~\ref{real_nests}). This is close to the region in Arabia Terra where mound-hosting craters are common \citep{Lewis2014, Bennett2016,Tanaka2000, Hynek2017}. Simulations were performed at four different times of year ($L_\mathrm{S} = 45^\circ$, 135$^\circ$, 225$^\circ$ and 315$^\circ$). Results are similar in all periods, and the results from $L_\mathrm{S} = 315^\circ$ are shown in Figure~\ref{real_atmos}. (See Movies S3-S4 in the supporting information for full diurnal results at each $L_\mathrm{S}$).

\begin{figure*}[t]
  \begin{center}
  \noindent\includegraphics[width=0.99\textwidth]{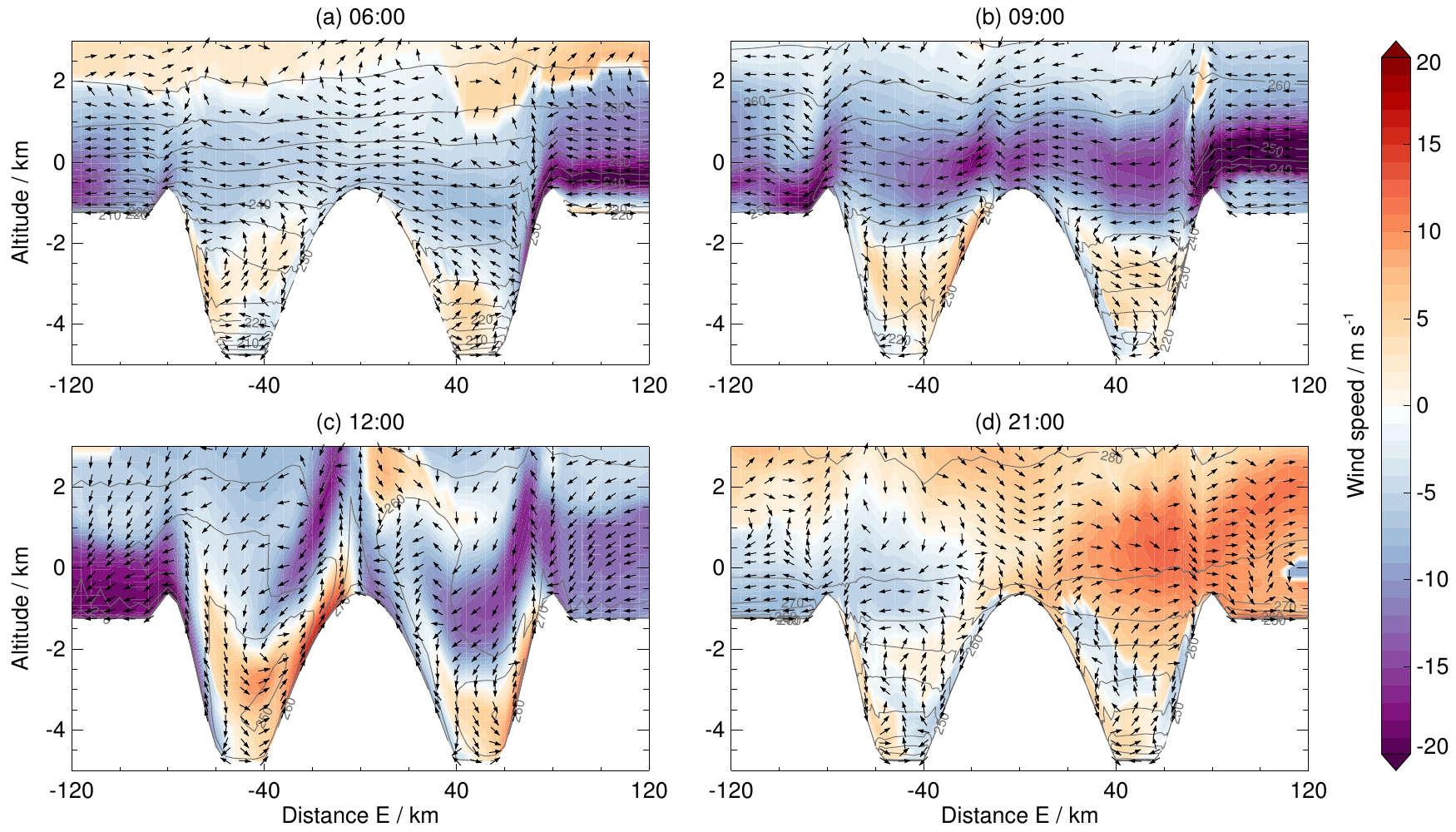}
  \caption{Wind speed (shaded), wind direction (arrows) and potential temperature (contours) at four different times from longitude-altitude slices through the center of the crater shown in Figure~\ref{real_atmos} in an east-west direction. Potential temperature is contoured at 5\,K intervals.}
  \label{real_atmos_circ}
  \end{center}
\end{figure*}

It is evident that the behavior is similar to the idealized cases, with downslope winds at night and strong upslope winds during the afternoon which increase in strength as they travel up the crater walls and mound flanks. The main difference is that the external wind field, which is strongest in the morning and afternoon, blowing from east to west (Figure~\ref{real_atmos}b,c), causes the stress field to be non-axisymmetric . The effect of this wind field is to increase the surface wind stress on the western crater wall and the leeward slope of the mound. Figure~\ref{real_atmos_circ} shows the circulation in a longitude-altitude plane taken across the center of the crater, with the times corresponding to those of Figure~\ref{real_atmos}.

At 06:00 the downslope wind is strongest on the eastern crater wall, as the slope is oriented in the same direction as the prevailing wind. By 09:00 the upslope flow over the mound flanks has developed, and is stronger than the flow on the crater walls. By 12:00 the upslope flows are fully developed. At the top of the mound there is convergence of the upslope flow, and the air is transported upwards away from the crater (Figure~\ref{real_atmos_circ}c), resulting in lower surface wind stresses at the top of the mound (Figure~\ref{real_atmos}c). At this time the upslope flow is strongest on the leeward slope of the mound. This is because the windward slope is affected by the prevailing wind, which, by the time it arrives at the mound, is traveling westward and downward. The subsiding component results in adiabatic warming, and thus the temperature contrast between the mound flank and the surrounding air is reduced (compare the potential temperature contours on either side of the mound in Figure~\ref{real_atmos_circ}c), limiting the strength of the upslope flow on the mound. The leeward slope is shielded from the prevailing easterly wind, and so the air temperatures are cooler, there is a larger temperature contrast between the mound flank and the surrounding air, and the upslope flow (and hence surface wind stress) can become stronger.

By contrast, \citet{Day2016} found that the mound in their wind tunnel experiments was preferentially eroded on the windward flank. However, as noted earlier, the size of their model crater means the mound is more likely to feel the direct effects of the wind, and be eroded. Additionally, such small models cannot take into account the changes in temperature experienced within real craters. Again, the mechanism inferred by \citet{Day2016} may occur in smaller diameter craters, while mounds in larger diameter craters may experience different erosional patterns. An example of such a case is Gale crater, where Mt.\ Sharp is offset in the opposite direction to the prevailing wind direction \citep{Bennett2016}, which is the behavior suggested by the erosion patterns in our simulations. Indeed, if erosion follows the surface wind stress field shown in Figure~\ref{real_atmos}c, then our work suggests an explanation for the `bat-wing' shape of Mt.\ Sharp.

\section{Discussion}

Our results show that winds on topographic slopes can potentially erode intra-crater sedimentary deposits to produce mounds. Mound evolution depends on the size of the host crater, with erosion in smaller craters resulting in mounds that are more squat, and erosion in larger craters resulting in steeper-sided mounds surrounded by moats. This behavior agrees with the mound morphologies in craters mapped by \citet{Bennett2016}. If craters are initially covered in sedimentary layers, more complex erosion patterns emerge, and can result in crescent-shaped moats with mounds joined partly to the crater rim. Large-scale winds blowing over large mound-hosting craters can result in the mound eroding more on the leeward side, with the center of the mound appearing to `march upwind' over time. This would result in a mound offset towards the direction of the prevailing wind, such as is observed for Mt.\ Sharp in Gale crater. Due to the strong day/night cycle of slope winds within canyons \citep[e.g.][]{Kite2016}, the results presented here may also apply to the formation of mounds within canyon systems such as Valles Marineris.

Implicit in these results is that saltation-abrasion is the landscape-modifying mechanism. We do not consider other processes that may have operated in a warmer or wetter environment, as erosion by liquid water has not been globally significant since the Late Noachian/Early Hesperian \citep{Golombek2006}. We assume detachment-limited erosion, i.e.\ that the timescale for weathering the sediment is much longer than the timescale to transport sediment out of the crater, and thus we do not follow the motion of individual particles. We know that small dust particles can remain suspended in the atmosphere of Mars in the present day, so it is likely that over time attrition will result in sedimentary particles becoming smaller, at which point they can be transported away from the crater in the daytime upslope winds. Larger abrading clasts may remain in the crater moat, as is evidenced in the Bagnold Dune Field in Gale crater \citep{Hobbs2010, Charles2017}. Behavior such as this might result in increased erosion of the moat, resulting in a positive feedback mechanism.

The simulations performed here are for atmospheric conditions relevant to present-day Mars, whereas much of the erosion of sedimentary mounds likely occurred billions of years ago \citep{Thomson2011, Palucis2016, Kite2017} when the atmosphere may have been much more dense \citep{Jakosky2017}. However, the main features noted here that are responsible for the erosion -- the upslope and downslope winds -- will still occur in a denser atmosphere. For example, slope winds are a common feature on Earth \citep[e.g.][]{Renfrew2006, Whiteman2010, Haiden2011, Munoz2013, Villagrasa2013, Lehner2016, Shapiro2016}. Indeed, the diurnal variation of temperature profiles within Meteor Crater in Arizona \citep{Whiteman2010} is similar to that in our simulations. However, in small craters like Meteor Crater, the strength of slope flows is limited due to the shallow depth, and so erosion is likely to be caused by smaller-scale features, such as those noted in large eddy simulations \citep{Day2016, Anderson2017}. Thus, the main features and processes noted here are still likely to occur in a denser Martian atmosphere, though the strength of the winds, and hence the potential erosion rates, are likely to differ.

It should also be noted that there are features of the circulation not modeled here, and which may potentially affect erosion over long timescales. For example, dust devil tracks have been observed in many craters \citep{Reiss2016}, and dust devils have been detected in-situ by rovers in Gusev and Gale crater \citep[e.g.][]{Greeley2006, Greeley2010, Moores2015, Kahanpaa2016, Steakley2016, Etxeberria2018}. However, simulations of crater circulations have shown that the boundary layer is suppressed within craters \citep[e.g.][]{Tyler2015, Rafkin2016}, which should limit the formation of dust devils in deep craters (indeed, fewer were detected in Gale crater compared to the shallower Gusev crater). Thus, while convective vortices have the ability to remove dust from the surface \citep[e.g.][]{Balme2006a, Balme2006b, Neakrase2016, Koester2017}, it is unlikely that dust devils contribute greatly to erosion rates within craters in the present-day. This may have been different in past climates, however \citep{Newman2005, Haberle2006}.

In the future, our ideas could be tested and refined by better constraints on erosion rates and patterns using crater counts \citep{Kite2017} and cosmogenic isotope exhumation-age dating \citep{Farley2014}.

\section{Conclusions}

While sedimentary mounds exist in craters of many different sizes, data \citep{Bennett2016} suggest that there is a tendency for intra-crater mounds completely encircled by moats to become more frequent as the crater diameter increases, hinting at a characteristic length scale (crater diameter) for encircling moats. We have performed mesoscale simulations considering craters 40, 80 and 160\,km in diameter, with depths extending to 3.5\,km, and a variety of mound and crater morphologies, to understand the formation of these sedimentary mounds.

\begin{enumerate}

\item Using a physically self-consistent numerical model, we find that mounds can form through wind erosion from craters surrounded by flat topography and filled with sediment. For a crater that is shallow, erosion will be fairly constant across the crater floor, resulting in an increase in the crater depth. As the depth increases to $\sim$2\,km, slope winds become more important, and result in increased erosion near the crater walls, forming a mound. However, if the sediment-filled crater is much deeper than $\sim$2\,km, the erosion near the crater walls reduces, and mound formation would either slow or stop completely.

\item Once a mound has formed, its evolution depends on the size of the host crater and its depth within the crater. For craters 40 and 80\,km in diameter, the surface wind stress distributions in the simulations (used as a proxy for erosion) suggest that mounds would erode more at their tops than at their bases, eventually becoming flatter. Conversely, mounds in the 160\,km crater would erode more at the sides and base than at the top, becoming thinner. This behavior is in agreement with observations: smaller craters tend to have proportionally shorter mounds. As mounds become more eroded and exist deeper in the crater, the weaker near-surface circulation reduces the surface wind stress, limiting the erosion. This may help to explain why mounds persist rather than being completely obliterated. 

\item In the case of a large-scale sedimentary layer covering the craters \citep[e.g.][]{Fergason2008, Bennett2016} the surface wind stress patterns in the simulations suggest that as the sedimentary layer recedes across a crater, it will erode more towards the edges of the crater, which could explain the appearance of some of the mounds that are still joined to the crater wall. 

\item When considering more realistic (GCM) meteorological boundary conditions, the main difference compared to the idealized simulations is the presence of a large-scale prevailing wind. The effect of this wind is to increase the surface wind stress values on the leeward side of the mound. The reason for this is that downwelling air on the windward side limits the strength of the daytime upslope flow. The leeward side experiences less downwelling air, and so the upslope wind can become stronger, increasing the surface wind stress and hence potential erosion. While most mounds are offset in the direction of the prevailing wind \citep{Bennett2016}, Mt.\ Sharp is offset in the opposite direction. The behavior in our simulations may offer an explanation for this offset, and for the `bat-wing' shape of Mt.\ Sharp.

\end{enumerate}

\section*{Acknowledgments}
We thank Mackenzie Day and an anonymous reviewer for their helpful comments which improved this paper. We thank Jasper Kok and Rob Sullivan for discussions of saltation on Mars, Daniel Tyler and Jeffrey Barnes for providing simulation results for benchmarking our model, Scot Rafkin for providing assistance with simulations, and the University of Chicago Research Computing Center. This work was funded in part by NASA grant NNX15AH998G. Model output is available to download from https://psd-repo.uchicago.edu/kite-lab/mesoscale\_crater\_data.

\end{document}